%% file: masonryPiers.tex
\documentclass[preprint,times,12pt,3p]{elsarticle}

\usepackage{amssymb}
\usepackage{amsmath}
\usepackage{fullpage}
\usepackage{graphicx}
\usepackage[colorlinks=true, linkcolor=blue, citecolor=blue]{hyperref}
\usepackage{multirow}
\usepackage{upgreek}
\usepackage{epstopdf}
\usepackage{color}
\usepackage{caption}
\usepackage{subcaption}

\newcommand{\subs}[1]{_{\mathrm{#1}}}   % subscript
\biboptions{sort&compress}

\journal{arXiv}

\begin{document}

\begin{frontmatter}

\author[ctu]{V. Ne\v{z}erka\corref{cor1}}
\ead{vaclav.nezerka@fsv.cvut.cz}
%\ead[url]{http://mech.fsv.cvut.cz/~nezerka/index.html}
\cortext[cor1]{Corresponding author}
\author[ctu]{J. Anto\v{s}}
\ead{jakub.antos@fsv.cvut.cz}
\author[ctu]{J. Lito\v{s}}
\ead{litos@fsv.cvut.cz}
\author[ctu]{P. Tes\'{a}rek}
\ead{tesarek@fsv.cvut.cz}
\author[ctu]{J. Zeman}
\ead{zemanj@cml.fsv.cvut.cz}

\address[ctu]{Faculty of Civil Engineering, Czech Technical University in Prague, Th\'{a}kurova~7, 166~29 Praha~6, Czech Republic}

\title{An Integrated Experimental-Numerical Study of the Performance of Lime-Based Mortars in Masonry Piers Under Eccentric Loading}

\begin{abstract}

Architectural conservation and repair are becoming increasingly important issues in many countries due to numerous prior improper interventions, including the use of inappropriate repair materials over time. As a result, the composition of repair masonry mortars is now being more frequently addressed in mortar research. Just recently, for example, it has become apparent that Portland cement mortars, extensively exploited as repair mortars over the past few decades, are not suitable for repair because of their chemical, physical, mechanical, and aesthetic incompatibilities with original materials. This paper focuses on the performance of various lime-based alternative materials intended for application in repairing historic structures when subjected to mechanical loading. Results of basic material tests indicate that the use of metakaolin as a pozzolanic additive produces mortars with superior strength and sufficiently low shrinkage. Moreover, mortar strength can be further enhanced by the addition of crushed brick fragments, which explains the longevity of Roman concretes rich in pozzolans and aggregates from crushed clay products such as tiles, pottery, or bricks. An integrated experimental-numerical approach was used to identify key mortar parameters influencing the load-bearing capacity of masonry piers subjected to a combination of compression and bending. The simulations indicate increased load-bearing capacities for masonry piers containing metakaolin-rich mortars with crushed brick fragments, as a result of their superior compressive strength.

\end{abstract}

\end{frontmatter}

\section{Introduction} \label{sec:introduction}

% architectural heritage - conservation and protection
Ancient structures embody the culture and stories of people, who built, used and lived in them. This charm attracts tourists to the sites with well-preserved cultural heritage, which in turn has an enormous positive impact on the economy of the region. From this reason, the conservation and restoration of architectural heritage is encouraged in the majority of countries. However, an inappropriate intervention can cause a huge harm, and therefore the authorities established numerous requirements on the procedures and materials used for the conservation and repairs.

% historic masonry structures and improper interventions
A vast number of ancient structures are made of masonry, being a traditional construction material that exhibits an extraordinary durability if an adequate maintenance is provided. Masonry bed joints are usually the weakest link and the deterioration and damage concentrates there. It has been established that the mortars used for repairs should be compatible with the original materials; serious damage to a number of historic masonry structures has been caused by an extensive use of Portland cement mortar over the past decades. The intention for its use was to avoid the inconveniences connected with the originally used lime-based mortars, such as slow setting, high shrinkage and low strength~\cite{Arizzi_2012}. However, the use of the Portland cement mortars has been reconsidered for their low plasticity, excessive brittleness and early stiffness gain~\cite{Veniale_2003, Callebaut_2001, Moropoulou_2005, Velosa_2009}. Moreover, the relatively high content of soluble salts that leach over time~\cite{Moropoulou_2005, Velosa_2009, Callebaut_2001} can severely damage the original masonry units because of large crystallization pressures~\cite{Seabra_2007, Sepulcre_2010} and produce anaesthetic layers on their surface.

% re-use of traditional pozzolans
The strict regulations with respect to the Portland cement use led to the exploitation of traditional additives to lime-based mortars, such as volcanic ash, burnt clay shale~\cite{Moropoulou_2004} or increasingly popular metakaolin~\cite{Velosa_2009}. These additives, known as \emph{pozzolans}, have been used since the ancient times in combination with lime to improve a moisture and free-thaw resistance of mortars~\cite{Slizkova_2009}, to increase their durability~\cite{Arizzi_2012, Velosa_2009} and also their mechanical strength~\cite{Papayianni_2006, Papayianni_2007}. The use of pozzolans is essential not only for bed joint mortar but also for rendering ones, because pure-lime mortars suffer from enormous shrinkage cracking that has a negative aesthetic impact and can even cause spalling of the facade surface layers~\cite{Wilk_2013}.

% use of crushed bricks
If there was no natural source of pozzolans available in the region, ancestors tried to find alternatives. Phoenicians were probably the first ones to add crushed clay products, such as burnt bricks, tiles or pieces of pottery, to the mortars in order to increase their durability and strength. Crushed bricks were often added to mortars used in load-bearing walls during the Roman Empire~\cite{Mallinson_1987} and Romans called the material \emph{cocciopesto}~\cite{Farci_2005}. Cocciopesto mortars were then extensively used from the early Hellenistic up to the Ottoman period in water-retaining structures to protect the walls from moisture, typically in baths, canals and aqueducts~\cite{Sbordoni_1981, Degryse_2002}. The brick dust was mainly used for rendering, while large pebbles up to 25~mm in diameter appeared mainly in masonry walls, arches and foundations~\cite{Baronio_1997}. However, our previous studies~\cite{Nezerka_2013_pastes, Nezerka_2014_nanoindentation, Nezerka_2016_model} revealed that the positive impact of ceramic fragments should not be attributed to the formation of hydration products due to limited reactivity, but rather to their compliance which limits shrinkage-induced cracking among aggregates and ensures a perfect bond with the surrounding matrix.

% introduce to the study and outline
The presented study was focused on the investigation of various mortars commonly used for repairs of cultural heritage and their structural performance through comprehensive experimental and numerical analyses. In particular, lime-based mortars with various additives and aggregates, introduced in Section~\ref{sec:materials}, were used in bed joints of masonry piers subjected to a combination of quasi-static compression and bending. The purpose of the experimental analysis, described in Section~\ref{sec:experimentalTesting}, was to study the failure modes and crack patterns using Digital Image Correlation (DIC), assess the structural performance of individual mortars, and verify the proposed material model used for the Finite Element (FE) predictions, presented in Section~\ref{sec:numericalSimulations}. The FE analysis was consequently utilized in Section~\ref{sec:caseStudy} to assess the key material parameters influencing the load-bearing capacity, and to study the failure modes of the masonry piers containing mortars with variable properties, subjected to a combination of compression and bending.

\section{Materials} \label{sec:materials}

Compared to historic limes, today's commercial ones are very pure, despite the very benevolent regulating standard EN 459-1 requiring the mass of CaO and MgO in the commonly used CL-90 lime hydrate higher than 90\%. However, the presence of impurities in historic limes mortars was not always harmful~\cite{Vejmelkova_2012}, since the content of silica (SiO$_2$) and alumina (Al$_2$O$_3$) was responsible for their hydraulic character~\cite{Lanas_2004}.

The inconveniences connected to the use of modern lime, such as limited binder strength, slow hardening, enormous shrinkage, and consequent cracking and poor cohesion between the mortar and surrounding masonry blocks~\cite{Wilk_2013} can be overcome by the use of reactive additives rich in aluminosilicates, such as metakaolin or Portland cement. While metakaolin has been generally accepted by the restoration community~\cite{Velosa_2009, Vejmelkova_2012}, the use of Portland cement is on decline and the authorities for cultural heritage in many countries prohibit its additions to repair mortars~\cite{Veniale_2003, Callebaut_2001, Sepulcre_2010}. According to a few studies, calcium-silicate-hydrates (CSH) and calcium-aluminum-silicate-hydrates (CASH) are the main hydrated phases formed at the room temperature after the pozzolanic reaction of metakaolin and Ca(OH)$_2$~\cite{Cizer_2009, DeSilva_1993, Rojas_2002}. The metakaolin presence in lime-based mortars results in an enhanced strength and durability~\cite{Nezerka_2013_pastes}, while the vapor transport properties are superior to the mortars containing Portland cement~\cite{Sepulcre_2010}.

Beside the addition of pozzolans, shrinkage can be efficiently reduced by increasing the content of inert aggregates, since the stiff inclusions restrain the volume changes of the surrounding matrix~\cite{Wilk_2013, Rougelot_2009}, which is more pronounced in the case of bigger inclusions~\cite{Stefanidou_2005}. However, large stiff pebbles are responsible for a formation of microcracks~\cite{Nezerka_2016_model}. that have a negative impact on the mortar integrity and reduce the mortar strength and stiffness~\cite{Arizzi_2012, Lanas_2004, Lanas_2003}. Moreover, the shrinkage-induced cracking of mortars poor in pozzolans, or containing unsuitable aggregates, limits their use as renderings because of their poor aesthetic performance~\cite{Mosquera_2006}.

Even though it is generally accepted that the presence of sand aggregates increases the resistance of mortars against mechanical loading, there is a threshold beyond which any addition of aggregates makes the mortar weaker due to excessive microcracking and loss of cohesion between the grains and the surrounding matrix~\cite{Arizzi_2012}. By experience, the~1~:~3 binder to aggregate volume ratio has been established as the most suitable for repair mortars, providing a reasonable strength, shrinkage and porosity.
%It is well known that the strength of cement-based mortars is reduced with the increasing porosity, e.g.~\cite{Neville_1996}, but in the case of lime-based mortars this relationship is not that clear~\cite{Lanas_2003}.
Based on the study by Stefanidou and Papayianni~\cite{Stefanidou_2005} it seems most favorable to use the sand of grain-size ranging between~0 and~4~mm, resulting in mortars of the highest strength.

Vitrivius, Roman author, architect and engineer, who lived in the first century BC, recommended in his \emph{Ten Books on Architecture} to add some portion of crushed bricks into mortars in order to increase their durability and strength. According to Silva et al.~\cite{Silva_2009}, the amorphous components of brick fragments, mainly represented by aluminosilicates, are able to react with lime and make the interfacial surface alkaline. The reaction products are supposed to give mortars a hydraulic character, and fill the voids and discontinuities in the thickness of about~20$\upmu$m from the interface between the crushed brick fragments and the surrounding matrix~\cite{Moropoulou_2002, Boke_2006}. However, such processes can take place only only if the ceramic clay is fired at appropriate temperatures between~600 and 900$^{\: \circ}$C~\cite{Elsen_2006}, and the mortar is hardening in a sufficiently wet environment~\cite{Baronio_1997_2} for a considerable amount of time~\cite{Moropoulou_2002}. Even if the reaction takes place, the reaction-rim thickness is very limited and does not have any significant impact on the mortar properties, as proven by the results of nanoindentation of ancient mortar samples in our previous work~\cite{Nezerka_2014_nanoindentation}. More importantly, the relatively compliant crushed brick fragments relieve the shrinkage-induced stresses and reduce the number of microcracks within the mortar matrix~\cite{Nezerka_2014_Brno, Nezerka_2016_model}.

Beside the positive impact of crushed brick fragments on the mechanical properties and durability of the cocciopesto mortars, the use of crushed bricks brings another benefit --- the use of waste by-products from ceramic plants leads to a cost reduction and production of a more sustainable material.

\subsection{Prepared and Tested Mortars}

For our study, we used a commonly available white air-slaked lime (CL90) \v{C}ertovy schody of a great purity (98.98\% of CaO + MgO), produced in the Czech Republic. The most frequent particle diameter found in the lime hydrate was equal to 15~$\upmu$m and its specific surface area, determined by the gas adsorption, was equal to 16.5~m$^2$/g. The finely ground burnt claystone metakaolin with a commercial name Mefisto~L05, produced by \v{C}esk\'{e} lupkov\'{e} z\'{a}vody Inc., Nov\'{e} Stra\v{s}ec\'{i}, Czech Republic, was chosen as the pozzolanic material. This additive is rich in SiO$_2$ (52.1~\%) and Al$_2$O$_3$ (43.4~\%). Portland cement CEM~I~42.5~R produced in Radot\'{i}n, the Czech Republic was used as an alternative to metakaolin. The selected Portland cement was rich in CaO (66~\%), SiO$_2$ (20~\%), Al$_2$O$_3$ (4~\%), Fe$_2$O$_3$ (3~\%), SO$_3$ (3~\%) and MgO (2~\%), as provided by XRF analysis~\cite{Nezerka_2013_pastes}.

\begin{figure}[ht]
\centering
\includegraphics[width=0.5\textwidth]{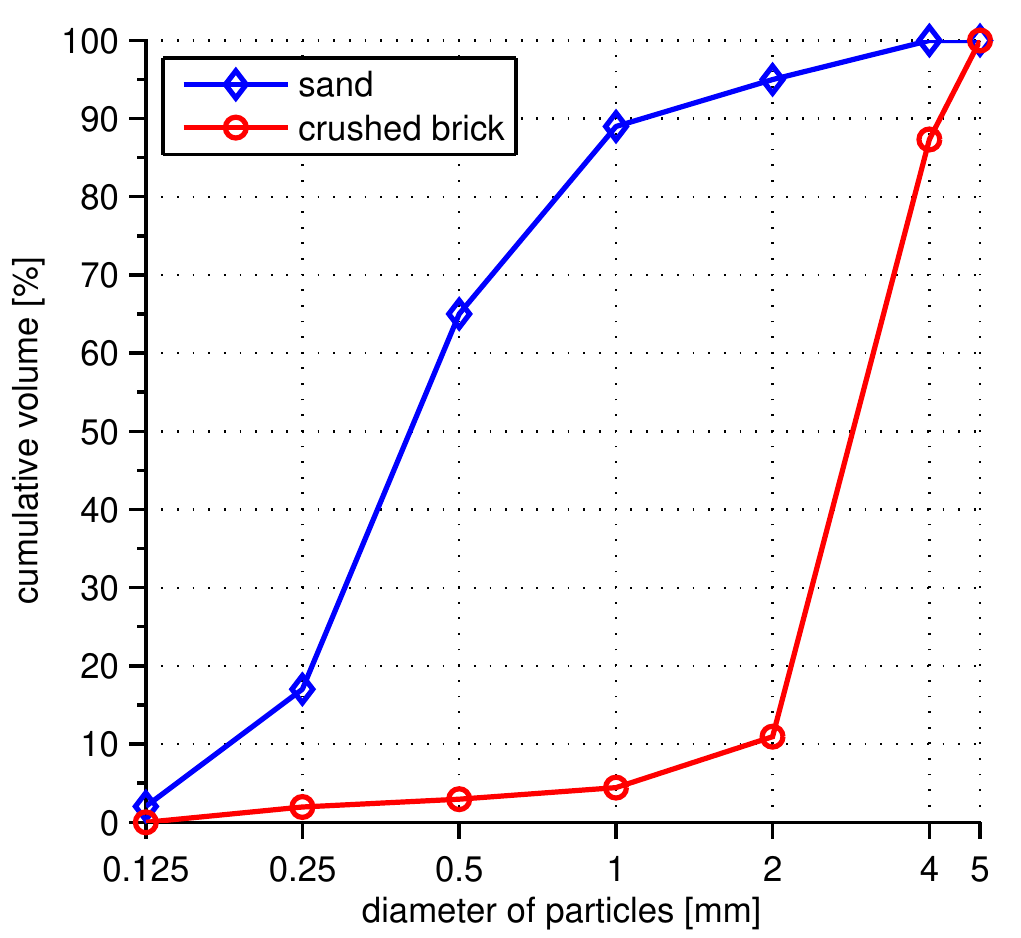}
\caption{Grading curves of sand and crushed brick aggregates.}
\label{fig:gradingCurves}
\end{figure}

Beside the investigation of metakaolin and Portland cement additions on the mechanical properties of lime-based mortars, the study was also focused on the influence of aggregate composition. River sand of grain size ranging between 0~and~4~mm from Z\'{a}lezlice was selected based on experience as the most suitable for the application as the bed joint mortar. The industrially produced crushed brick fragments of the grain-size 2--5 mm, from a brick plant Bratronice, the Czech Republic, were chosen based on results of previous studies~\cite{Nezerka_2014_EAN2013} and experience of authors acquired by analyses of ancient mortar samples~\cite{Baronio_1997, Moropoulou_1995, Moropoulou_2002}. The grain size distribution of the sand and crushed bricks aggregates, obtained by a sieve analysis, is presented in Figure~\ref{fig:gradingCurves}.

The mass ratio of lime and metakaolin~/~Portland cement was equal to~7:3 in all mortars. The amount of water was adjusted so that the fresh mortars fulfilled the workability slump test in accordance with~\v{C}SN~EN~1015-3 and the mortar cone expansion reached~13.5$\pm$0.3~cm. Such consistency ensured a sufficient workability while keeping the water to binder ratio (w/b) as low as possible to avoid shrinkage cracking. The amount of aggregates was designed based on our experience, previous studies~\cite{Arizzi_2012, Lanas_2004, Stefanidou_2005} and results of micromechanical modeling~\cite{Nezerka_2016_model} towards high strength and acceptable shrinkage. The composition of the tested mortars is summarized in Table~\ref{tab:mortarsComposition}.

\begin{table}[ht]
 \caption{Mass ratios of constituents in the tested mortars and their shrinkage after 90 days of hardening; PC and CB abbreviations stand for Portland cement and crushed bricks, respectively.}
 \label{tab:mortarsComposition}
 \centering
\begin{tabular}{l c c c c c c c c}
  \hline
  \multirow{2}{*}{mix}  & \multicolumn{3}{c}{binder} & & \multicolumn{2}{c}{aggregate}   & water / dry mass  & 90-days\\[1.5pt] \cline{2-4}\cline{6-7}
            & lime  & PC & metakaolin   & & sand & CB           & (water / binder) & shrinkage \\[1.5pt] \hline
  LC-S      & 0.7           & 0.3         & --       & & 3    & --                       & 0.175 (0.704)    & 0.71~\% \\
  LMK-S     & 0.7           & --          & 0.3      & & 3    & --                       & 0.180 (0.714)    & 0.83~\%\\
  LMK-SCB   & 0.7           & --          & 0.3      & & 1    & 1.5                      & 0.250 (0.875)    & 0.64~\%\\
  L-SCB     & 1.0           & --          & 0.3      & & 1    & 1.5                      & 0.320 (0.940)    & 1.10~\%\\ \hline
\end{tabular}
\end{table}

The crushed bricks aggregate retains more water than sand (see the water / dry mass ratio records in Table~\ref{tab:mortarsComposition}). Based on such finding, we conjecture that the presence of water-retaining crushed bricks can promote the hydraulic reactions within the binder, and increase mortar strength and stiffness.

\section{Experimental Testing} \label{sec:experimentalTesting}

The experimental testing consisted of two stages --- first the individual components, i.e. the mortars and masonry units, were subjected to series of three-point bending and compression tests in order to acquire the data necessary for the calibration of the FE model. The second stage involved a full-scale compression test of masonry piers.

\subsection{Acquisition of Basic Material Parameters} \label{sec:parametersAcquisition}

The basic material parameters, describing the mechanical behavior of mortars and bricks, were obtained from results of three-point bending and uniaxial compression tests, carried out according to EN~1015-11. To that purpose six control specimens were cast into 160~$\times$~40~$\times$~40~mm prismatic molds, compacted using shaking table to get rid of excessive air bubbles, and removed after 48~hours. Curing was executed at the temperature of 20$\pm$1$^{\: \circ}$C and relative humidity ranging between 60~and~80~\%.

Common fired clay bricks with dimensions 290~$\times$~140~$\times$~65~mm, produced in the brick plant \v{S}t\v{e}rboholy, the Czech Republic, were used as the masonry units for the construction of the tested masonry piers. In order to obtain the basic mechanical parameters, six 140~$\times$~40~$\times$~40~mm prisms were cut off the bricks and subjected to the three-point bending and compression tests. The same procedure was adopted for the mortar specimens.

\begin{figure}[ht]
\centering
\includegraphics[width=0.95\textwidth]{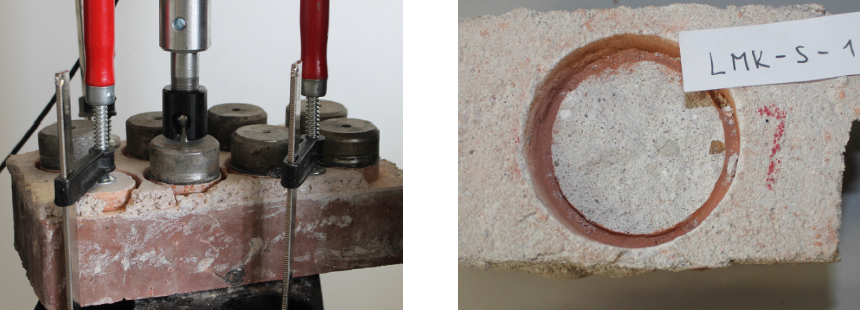}
\caption{Interface strength pull-out testing: experiment set-up (left) and the failure plane located within the mortar layer (right).}
\label{fig:interfaceTesting}
\end{figure}

Beside the bending and uniaxial compression tests, the dynamic Young's modulus was assessed by the resonance method on 90-day old samples, and the tensile strength of the interface between mortars and bricks was evaluated based on series of pull-out tests carried out in accordance with EN 1015-21. The tests revealed that the interface was not the weakest link in the case of all tested mortars, since the failure plane was not located at the brick boundary, Figure~\ref{fig:interfaceTesting}. The satisfactory interface strength is attributed to the relatively big roughness of the bricks and suitable workability of the fresh mortars.

\subsubsection{Resonance Method}

The non-destructive resonance method was utilized to assess the the dynamic Young's modulus, $E_{\mathrm{dyn}}$, of both, mortars and brick. Such approach was chosen to overcome the inconvenience connected to the measurement of the static Young's modulus arising from the load-dependent compliance of the loading frame or improper attachment of strain-gauges. According to Malaikah et al.~\cite{Malaikah_2004} the ratio between static and dynamic moduli measured on the same material should be within the range between 0.9 and 1.1.

The dynamic Young's modulus measurement is based on the equation for a longitudinal vibration of the beam with a continuously distributed mass and free-free boundary condition, following
\begin{equation}
	E_{\mathrm{dyn}}=\frac{4Lmf_{\mathrm{I}}^2}{bt},
\end{equation}
where $L$ is the length of the specimen [m], $m$ is the mass of the specimen [kg], $f_{\mathrm{I}}$ is the fundamental longitudinal resonant frequency of the specimen [Hz], $b$ is the width [m] and $t$ is the thickness of the specimen [m]. For detailed information on the procedure of the dynamic Young's modulus assessment the reader is referred e.g.~to~\cite{Radovic_2004}.

The obtained values of the dynamic Young's modulus were used for re-scaling the displacements provided by the gauge attached to the loading frame, when evaluating the load-displacement diagrams provided by the three-point bending and uniaxial compression tests.

\subsubsection{Three-Point Bending}

The displacement-controlled three-point bending tests were performed on unnotched 160~$\times$ 40~$\times$ 40~mm simply supported beams with distance between supports equal to 100~mm in order to obtain the tensile failure-related material parameters. The loading was introduced in the midspan at the rate of 0.025~mm/min in order to capture the descending part of the load-displacement diagram and monitored using MTS Alliance RT 30~kN load cell.

\subsubsection{Uniaxial Compression}

The uniaxial compression test was carried out on cubic 40~$\times$~40~$\times$~40~mm specimens, cut off the halves of the cracked beams from the three-point bending test, using the same device as for the three-point bending. A uniform contact and force distribution was accomplished by loading the flat lateral sides, being in contact with mold during preparation of the specimens. The loading was displacement-controlled at the rate of 0.3~mm/min.

\subsection{Testing of Masonry Piers}

\begin{figure}[ht]
\centering
    \def\svgwidth{0.45\linewidth}
    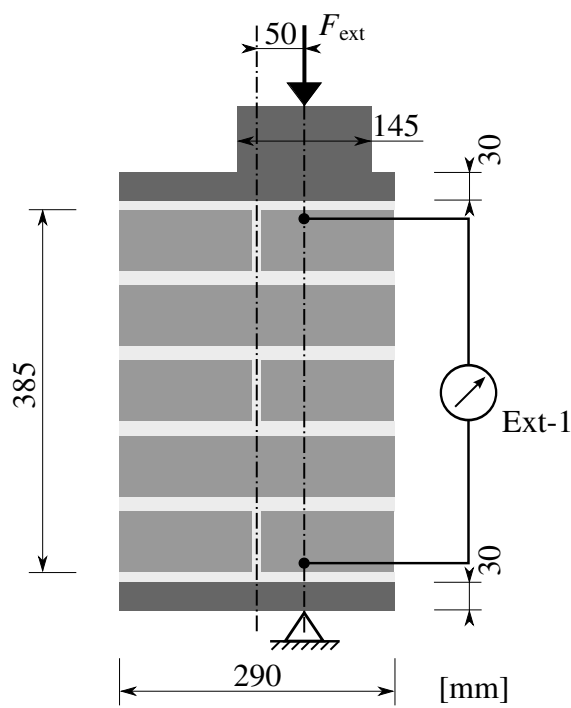
\caption{Loading of the tested piers and placement of virtual extensometers.}
\label{fig:schemePiers}
\end{figure}

The geometry of the tested masonry piers of a square cross-section is described in the scheme provided in Figure~\ref{fig:schemePiers}. The piers were subjected to a displacement controlled quasi-static compression with an eccentricity to introduce a combination of bending and compression. The geometry, loading, boundary conditions, and material of the masonry blocks (clay bricks) were the same for all tested piers, because the study was focused on the influence of the bed joint mortar. The bed joint thickness was equal to 15~mm, while the vertical joints were 10~mm thick, and the bricks were arranged in five layers to make a proper bond.

The loading of the piers was accomplished using a steel loading frame with a hydraulic actuator of 1~MN loading capacity. The steel slabs ensured a uniform distribution of stresses from the loading and a joint assembly allowed a rotation of the pier ends in all directions. The test was displacement-controlled with a loading rate of 1~mm/min in order to capture the descending part of the force-displacement diagram after the loss of pier integrity.

\subsection{DIC}

Extensometers and strain-gauges, conventionally used to measure displacements and deformation at pre-determined locations, cannot provide information about strain localization phenomena, such as cracking. The full-field DIC measurement allows to track the strain-field over the region of interest and capture the damage initiation and its progression until the complete loss of structural integrity. This non-contact method was first mentioned in papers by Yamaguchi~\cite{Yamaguchi_1981}, followed by Peters and Ranson~\cite{Peters_1982}, and their pioneering work established the basic principles.

\begin{figure}[htp]
\begin{minipage}[b]{0.45\linewidth}
\centering
\includegraphics[width=\textwidth]{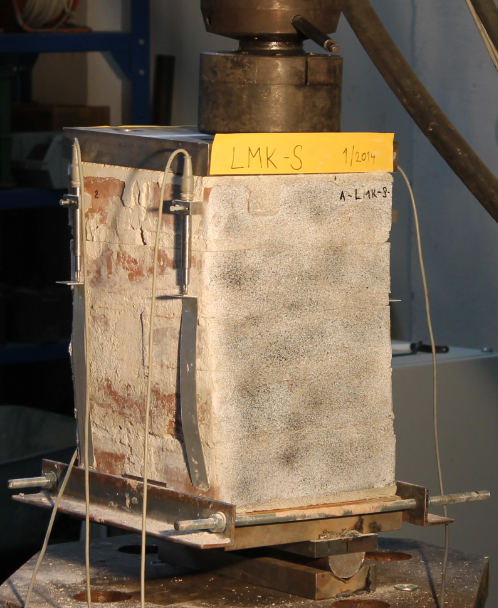}
\caption{Tested masonry pier subjected to the combination of compression and bending, introduced via eccentric loading and supporting.}
\label{fig:testedPier_EC}
\end{minipage}
\hspace{0.5cm}
\begin{minipage}[b]{0.45\linewidth}
\centering
\includegraphics[width=\textwidth]{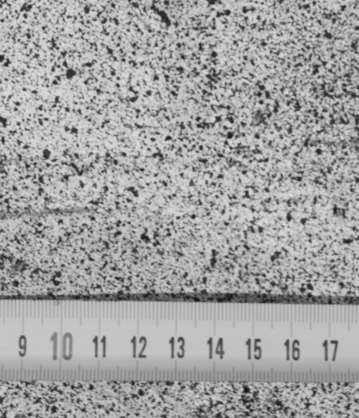}
\caption{Black and white stochastic pattern applied onto the pier surface for the purposes of DIC; the main scale units represent centimeters.}
\label{fig:testedPier_pattern}
\end{minipage}
\end{figure}

DIC is based on tracking the in-plane deformation and displacement of a surface texture or a stochastic pattern applied artificially on a sample, Figure~\ref{fig:testedPier_pattern}). A subset of gray-scale pixels in the reference image (representing an initial state) is matched to a subset with the best correlation within the deformed surface and its displacement and deformation are consequently evaluated. Using DIC, the deformation is not averaged over the strain-gauge length, because the averaging of strains depends purely on the camera resolution. Moreover, the technique is not limited to the measurement of small strains and the surface can be relatively rough as in the case of masonry, where strain gauges usually fail. The studies focused on the evaluation of errors in the field of displacements and deformations obtained by DIC, e.g.~\cite{Lava_2009, Bornert_2009}, demonstrate a relatively good accuracy of the method compared to conventional measurement techniques. A portion of the data can be lost if a spalling of the surface layer occurs, which easily happens in the case of quasi-brittle materials subjected to extreme loading, limiting the analysis of the post-peak behavior.

The optical monitoring of masonry piers was accomplished using high-definition Canon~70D camera taking pictures at 5-seconds time intervals, set equally for all tested piers, to yield on average 210 images documenting the deformation of a single pier until its complete failure. The light sensitivity ISO parameter was manually set to 100, since a powerful artificial lighting was available, Figure~\ref{fig:testedPier_EC}. The perfect illumination allowed the short exposure time (1/125~sec) and the low aperture number set to~$f$/8, which was kept constant for all images in the series. In order to minimize the effect of lens distortion~\cite{Pan_2013}, the distance between the camera and the observed surface was approximately 1~m, and the focal length (zoom) was set to 55~mm. The resulting DIC real scale resolution equal to~0.202 mm/pixel was computationally feasible while preserving the required precision.

\begin{figure}[htp]
\centering
   \begin{subfigure}{0.43\linewidth} \centering
     \includegraphics[width=\textwidth]{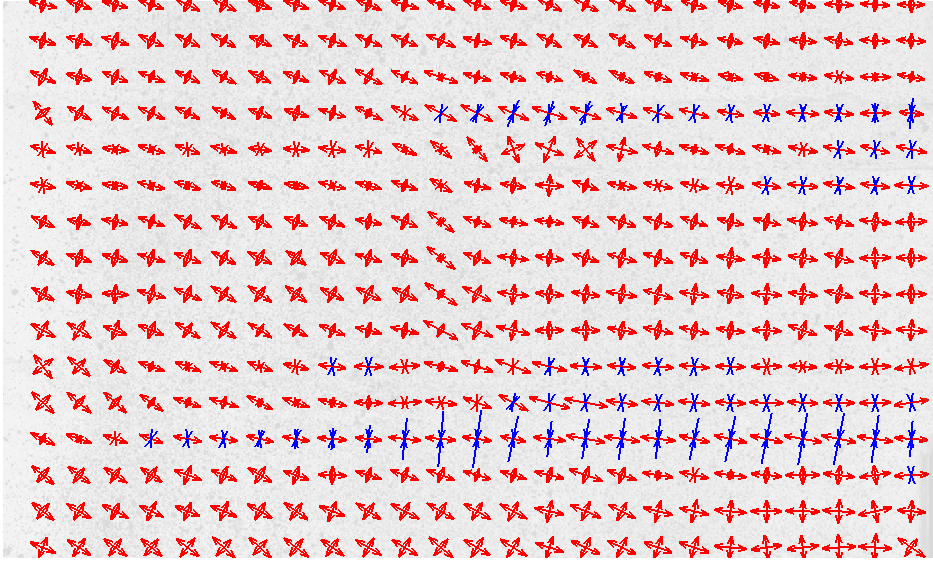}
   \end{subfigure}
   \hspace{0.1\linewidth}
   \begin{subfigure}{0.43\linewidth} \centering
     \includegraphics[width=\textwidth]{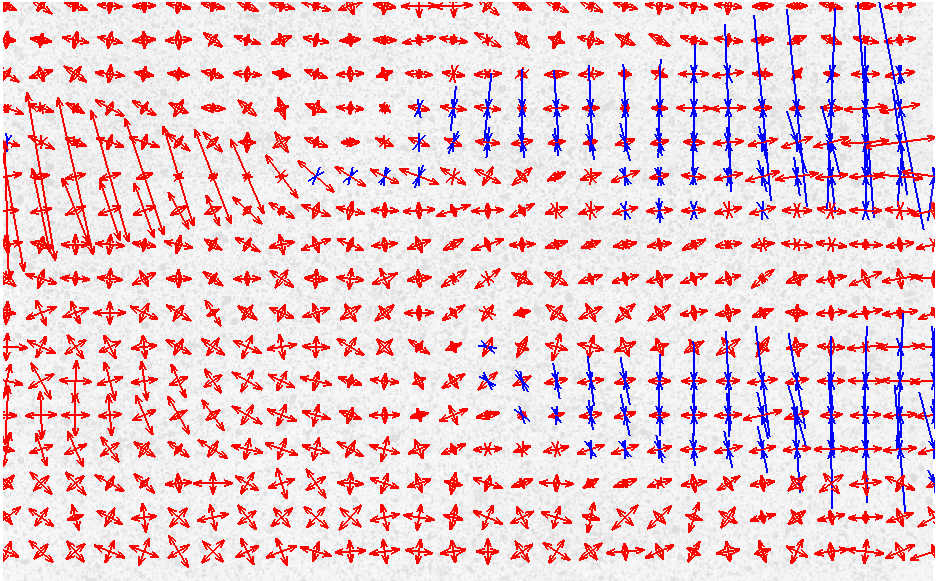}
   \end{subfigure}
\caption{Direction and relative magnitude of principal strains on the surface of masonry piers: stiff LMK-SCB mortar joints (left) and compliant L-SCB joints (right); red and blue arrows indicate principal tension and compression, respectively.}
\label{fig:arrowsPrincipalStrain}
\end{figure}

The displacement and strain fields were evaluated using an open source DIC software Ncorr~\cite{Blaber_2015}, and post-processing of the results was accomplished using Ncorr\_post tool~\cite{Ncorr_post}, both operating in MATLAB environment. The DIC results were used for the validation of the numerical model introduced next in Section~\ref{sec:numericalSimulations} and for capturing the strain localization. Figure~\ref{fig:arrowsPrincipalStrain} clearly demonstrates the difference between strain localization in bed joints in the case of a compliant (L-SCB) and a stiff mortar (LMK-SCB).

\section{Numerical Simulations} \label{sec:numericalSimulations}

The traditional design of masonry structures, based on rules-of-thumb, has been replaced by the numerical approach to address their complex failure mechanisms~\cite{Lourenco_2002}. Our FE simulations were employed to find the optimum properties of the bed-joint mortars to be used in masonry piers subjected to a combination of compression and bending, which represents a typical loading of masonry walls, columns, and vaults. Mortar shrinkage was neglected in the numerical simulations. The verification process was accomplished by comparing the numerical predictions with the experimentally obtained data; DIC allowed to compare the predicted and observed failure modes, and easily construct the force-displacement diagrams. 

In order to provide reliable predictions, the FE model must be supplied with a proper constitutive model. Based on experimental observations of masonry failure modes, damage-plastic models seem to provide the best description of both components, bricks and mortar. In the studies by Wawrzynek and Cincio~\cite{Wawrzynek_2005} and Zhou~et~al.~\cite{Zhang_2012} the isotropic damage-plastic models were able to accurately describe the response of masonry to mechanical loading even in non-linear domain after plasticity and cracking took place.

In our simulations we used 15k (uniaxial compression test specimen), 30k (three-point bending test specimen), and 80k (masonry pier) linear tetrahedral elements and employed a phenomenological damage-plastic constitutive model proposed by Jir\'{a}sek and Grassl~\cite{Grassl_2006}, implemented in the open-source FE package OOFEM~\cite{Patzak_2001}. The model was developed for the predictions of concrete failure subjected to general triaxial stress, and verified through a comprehensive experimental analysis.

The mesh generation was accomplished using Salome open-source generic platform and the post-processing of results and plotting the force-displacement diagrams was done using MATLAB scripts, while the graphical output was prepared in Paraview software.

\subsection{Material Model}
The material model~\cite{Grassl_2006} chosen to represent all components of the investigated masonry piers, i.e. studied mortars and bricks, considers the stress-strain law in the form of
\begin{equation}
    \boldsymbol{\sigma}=(1-\omega)\mathbf{D}_{\mathrm{e}}:(\boldsymbol{\varepsilon-\varepsilon}_{\mathrm{p}}),
\end{equation}
where $\boldsymbol{\sigma}$ is the stress tensor, $\omega$ is a scalar damage parameter (damage is assumed to be isotropic) and $\mathbf{D}_{\mathrm{e}}$ is the elastic stiffness tensor. In contrast to pure damage models the damage evolution is not driven by the total strain, $\boldsymbol{\varepsilon}$, but it is linked to the evolution of elastic strain, $\boldsymbol{\varepsilon}-\boldsymbol{\varepsilon}_{\mathrm{p}}$. The plastic part of the model formulation consists of a three-invariant yield condition, non-associated flow rule and a pressure-dependent hardening law. The mesh-independent response in the post-peak regime was achieved by the crack band approach~\cite{Bazant_1983,Jirasek_2012}.

\subsection{Identification of Material Model Parameters} \label{sec:inverseAnalysis}

\begin{figure}[htp]
\centering
   \begin{subfigure}{0.53\linewidth} \centering
     \includegraphics[width=\textwidth]{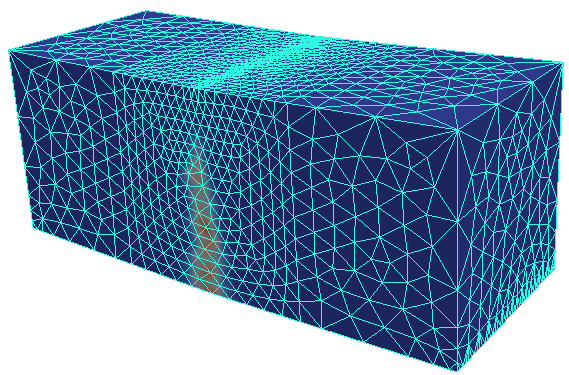}
   \end{subfigure}
   \hspace{0.1\linewidth}
   \begin{subfigure}{0.35\linewidth} \centering
     \includegraphics[width=\textwidth]{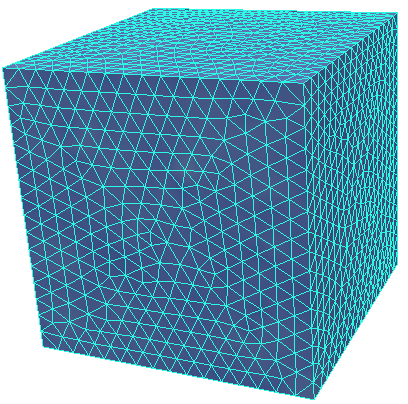}
   \end{subfigure}
\caption{FE model of specimens for three-point bending (left) and uniaxial compression (right) tests.}
\label{fig:calibrationSpecimens}
\end{figure}

The independent material model parameters to represent individual mortars and clay bricks were adjusted through an inverse analysis by reproducing the experimentally obtained data. First, the Young's modulus, $E$, and parameters to influence the flexural strength (tensile strength, $f_{\mathrm{t}}$, and fracture energy, $G_{\mathrm{f}}$) were adjusted to reproduce the results of the three-point bending tests, Figure~\ref{fig:calibrationSpecimens}.

The material model parameters to characterize the response to compression (compressive strength, $f_{\mathrm{c}}$, softening and hardening parameters, $A_{\mathrm{soft}}$ and $A_{\mathrm{hard}}$ --- see~\cite{Grassl_2006} for details) were calibrated by fitting the results from uniaxial compression tests. The Poisson's ratio was selected based on a literature study~\cite{Drdacky_2003, Vorel_2012, Nezerka_2012_AP} as~0.2 to represent both, mortars and bricks. The influence of Poisson's ratio, $\nu$, representing individual components was also analyzed to confirm its negligible impact on results when considered within reasonable bounds between~0.1 and~0.3, as reported in literature, e.g.~\cite{Corinaldesi_2009, Sandoval_2012}.

\begin{figure}[htp]
\centering
   \begin{subfigure}{0.43\linewidth} \centering
     \includegraphics[width=\textwidth]{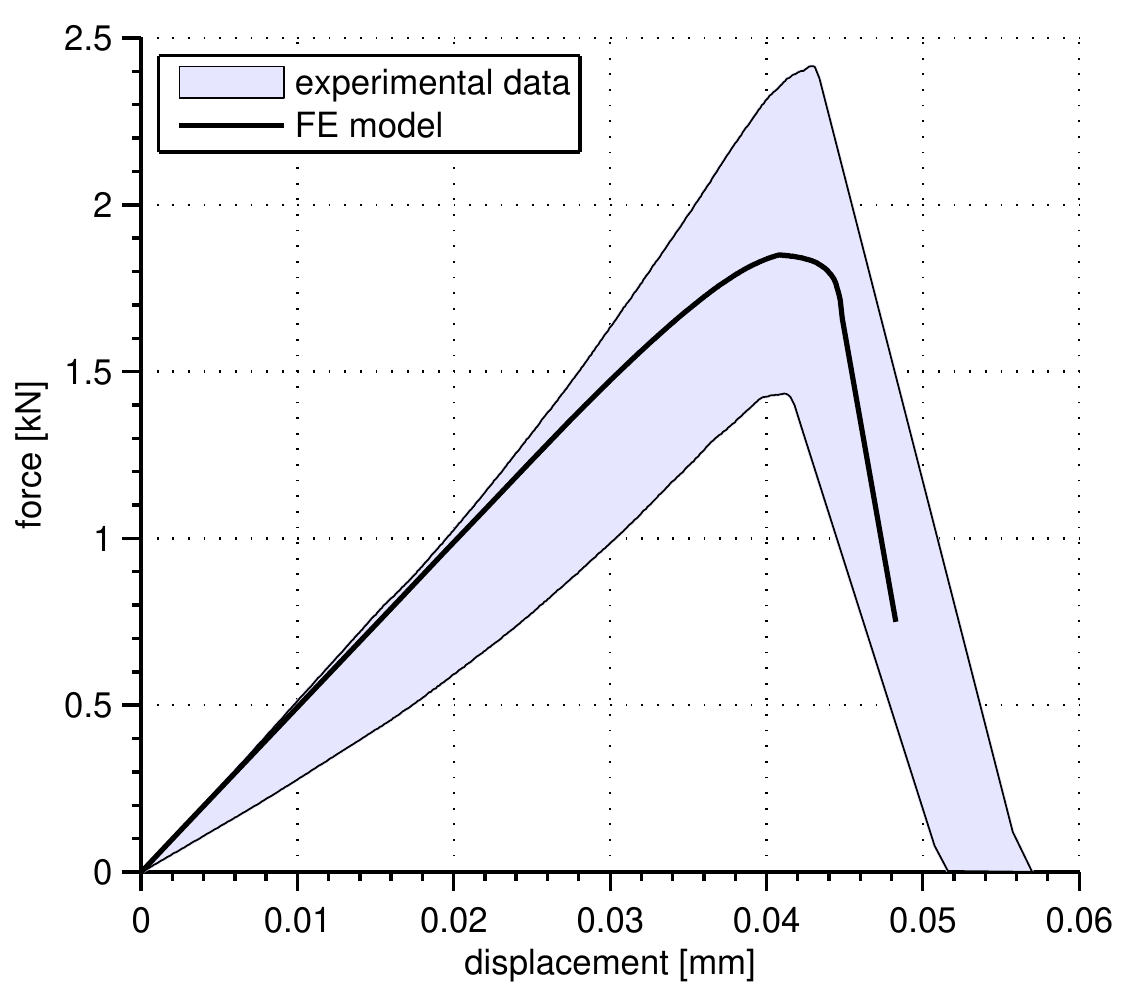}
   \end{subfigure}
   \hspace{0.1\linewidth}
   \begin{subfigure}{0.43\linewidth} \centering
     \includegraphics[width=\textwidth]{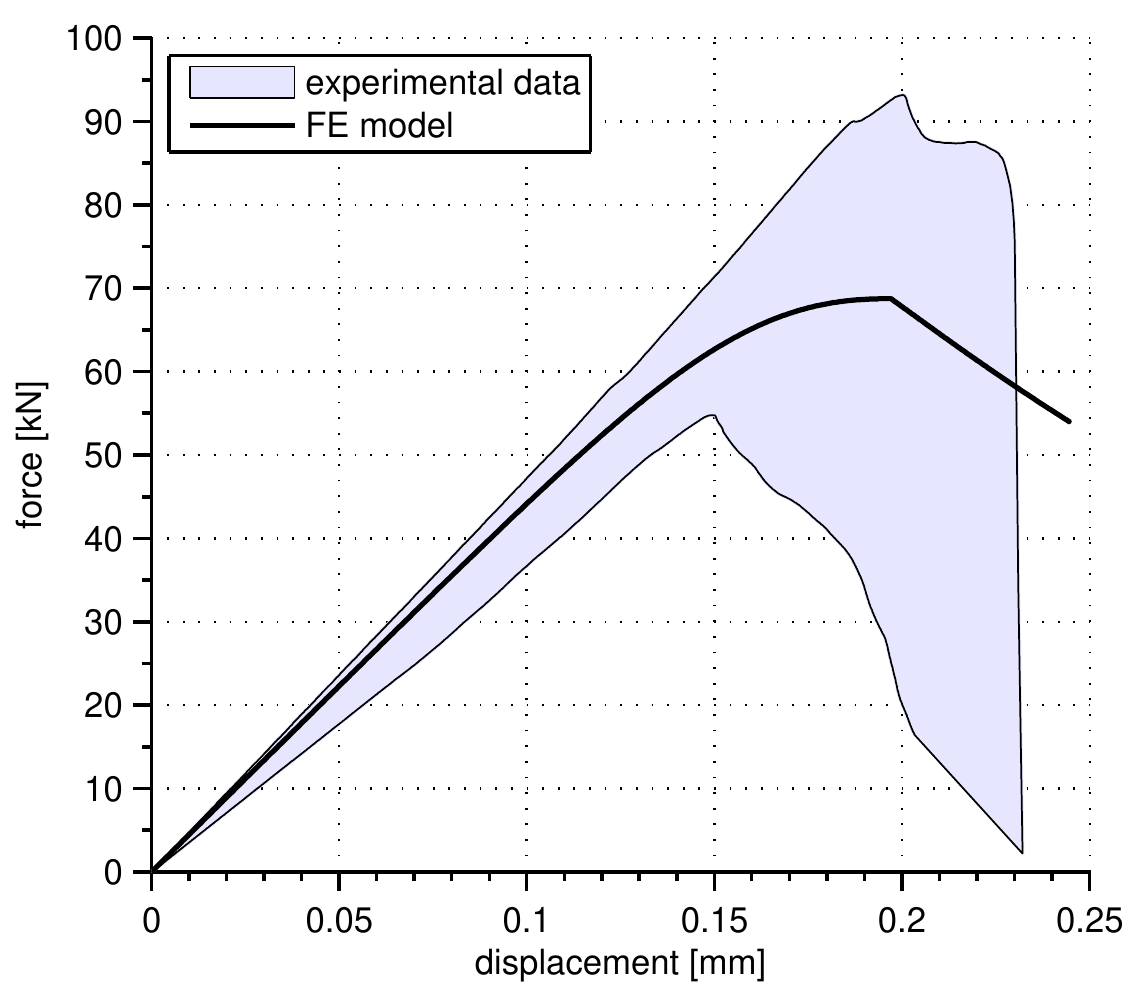}
   \end{subfigure}
\caption{Fit of the material model input parameters representing bricks to reproduce the results of three-point bending (left) and uniaxial compression (right) tests.}
\label{fig:calibrationBrick}
\end{figure}

\begin{figure}[htp]
\centering
\includegraphics[width=0.9\textwidth]{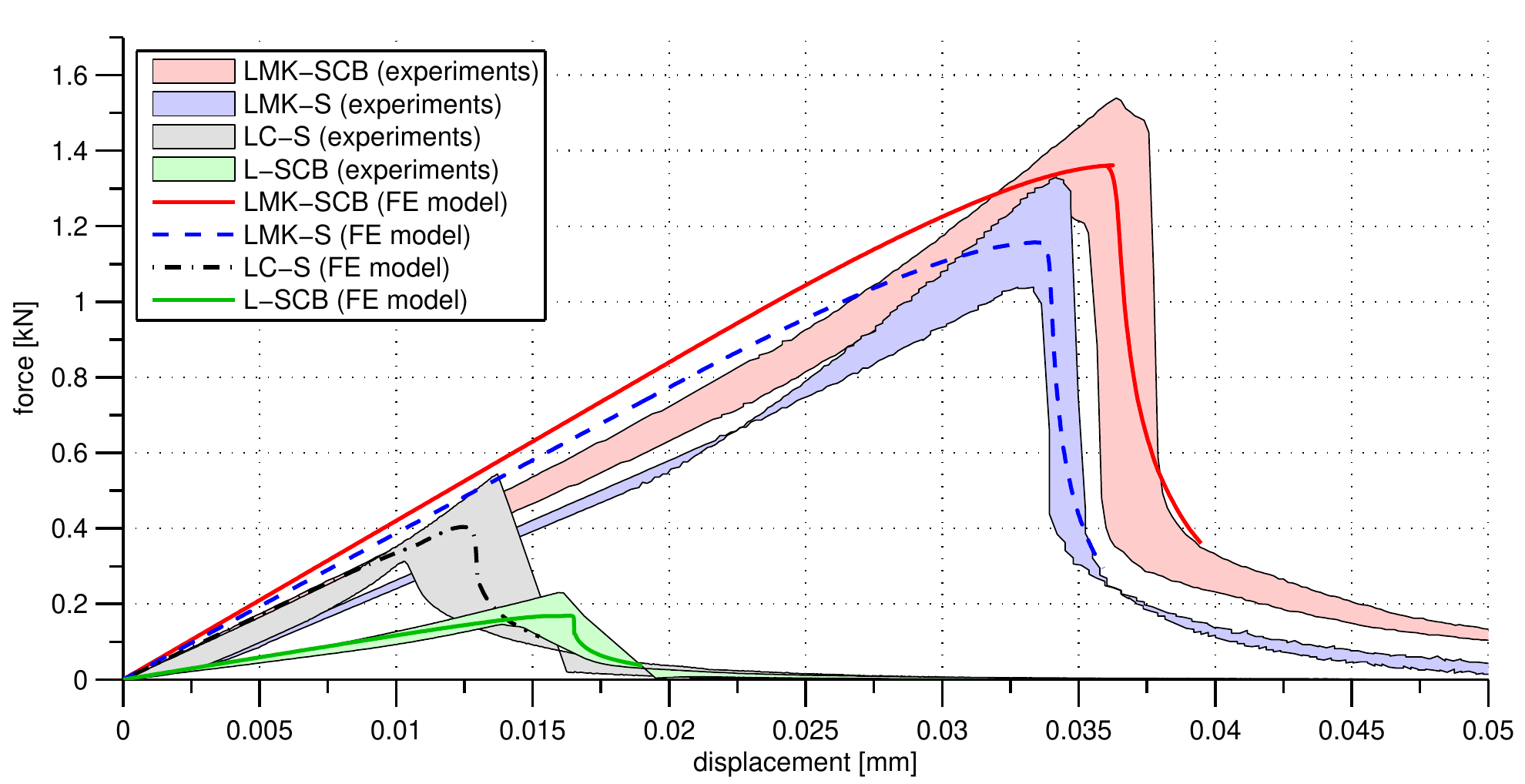}
\caption{Fit of the material model input parameters representing the individual tested mortars to reproduce the results of three-point bending tests.}
\label{fig:calibrationMortarsBending}
\end{figure}

\begin{figure}[htp]
\centering
\includegraphics[width=0.9\textwidth]{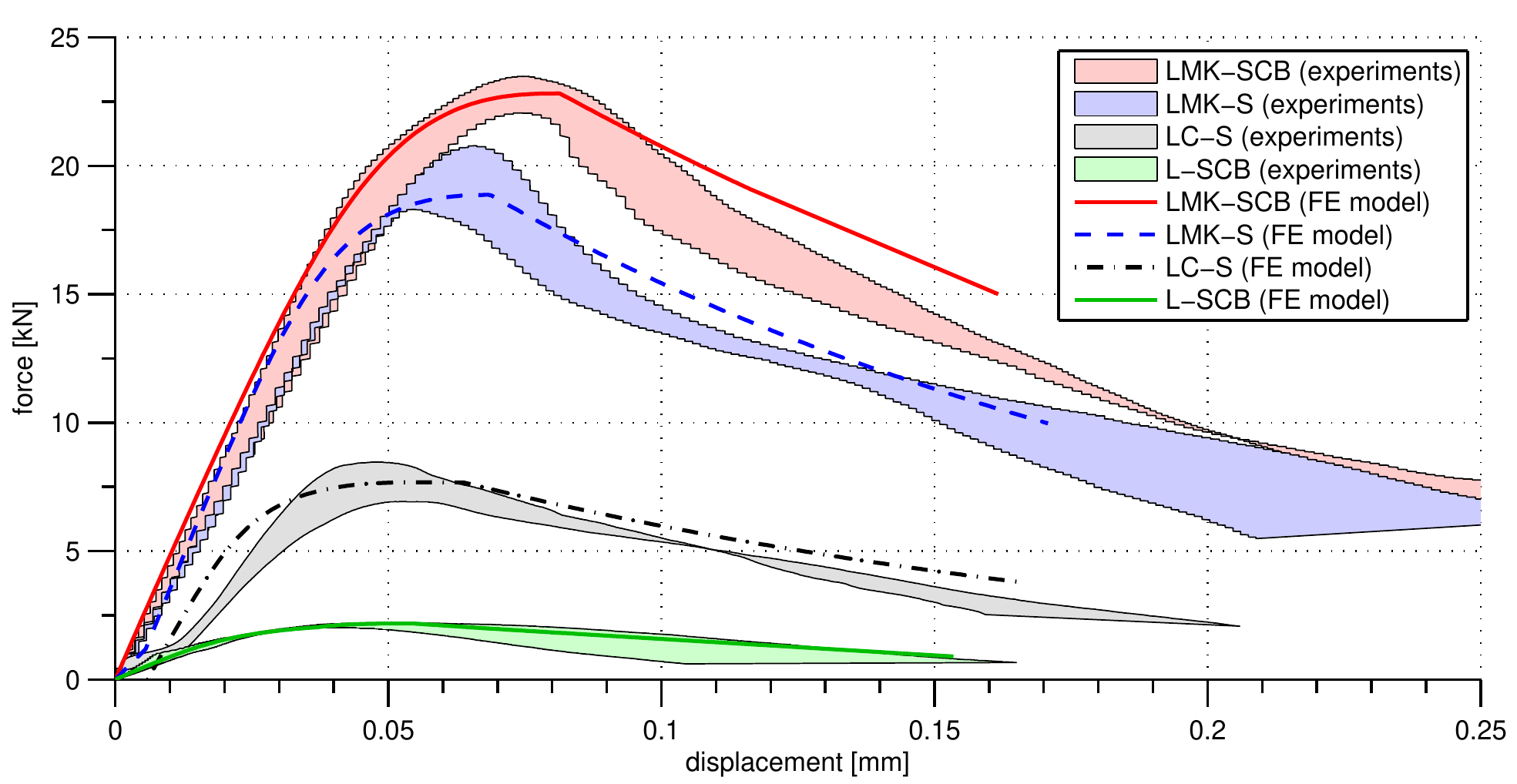}
\caption{Fit of the material model input parameters representing the individual tested mortars to reproduce the results of uniaxial compression tests.}
\label{fig:calibrationMortarsCompression}
\end{figure}

Even though the nature of the chosen material model did not allow to accurately reproduce the entire experimentally obtained load-displacement path, the results are, with respect to the large scatter of the experimental data (indicated by the shaded area in Figures~\ref{fig:calibrationBrick}, \ref{fig:calibrationMortarsBending}, and~\ref{fig:calibrationMortarsCompression}), considered satisfactory. The summary of the material model input parameters to represent individual mortars and bricks are summarized in Table~\ref{tab:materialProperties}.

%The shrinkage of mortars was neglected in the numerical simulations. However, the high shrinkage rate causing cracking of mortars poor in pozzolans and aggregates can be responsible for an excessive cracking~\cite{Nezerka_2013_pastes, Wilk_2013}, resulting in poor mechanical and aesthetic performance~\cite{Mosquera_2006}. Independent of mechanical properties, this inconvenience limits the use of pure lime mortars in heavily loaded elements or as renderings.

\begin{table}[ht]
 \caption{Key material input properties of bricks and the tested mortars; $E$, $\nu$, $f\subs{c}$, $f\subs{t}$ and $G\subs{f}$ denote the Young's modulus, Poisson's ratio, compressive strength, tensile strength and fracture energy, respectively.}
 \label{tab:materialProperties}
 \centering
 \renewcommand{\arraystretch}{1.2}
 \begin{tabular}{c c c c c c}
  \hline
  material       & $E$     & $\nu$  & $f\subs{c}$ & $f\subs{t}$ & $G\subs{f}$  \\
                 & [GPa]   & [-]    & [MPa]       & [MPa]       & [J/m$^2$]    \\  \hline
  bricks         & 14.0    & 0.20   & 30.0        & 2.70        & 30.0         \\
  LC-S           & 7.10    & 0.20   & 4.80        & 0.55        & 3.20         \\
  LMK-S          & 7.40    & 0.20   & 11.3        & 1.53        & 13.5         \\
  LMK-SCB        & 10.0    & 0.20   & 14.7        & 1.92        & 26.0         \\
  L-SCB          & 2.45    & 0.20   & 1.30        & 0.26        & 1.55         \\
  \hline
 \end{tabular}
\end{table}

\subsection{Numerical Simulations of Masonry Pier Failure} \label{sec:numericalSimulationsOfPiers}

\begin{figure}[htp]
\centering
\includegraphics[width=0.43\textwidth]{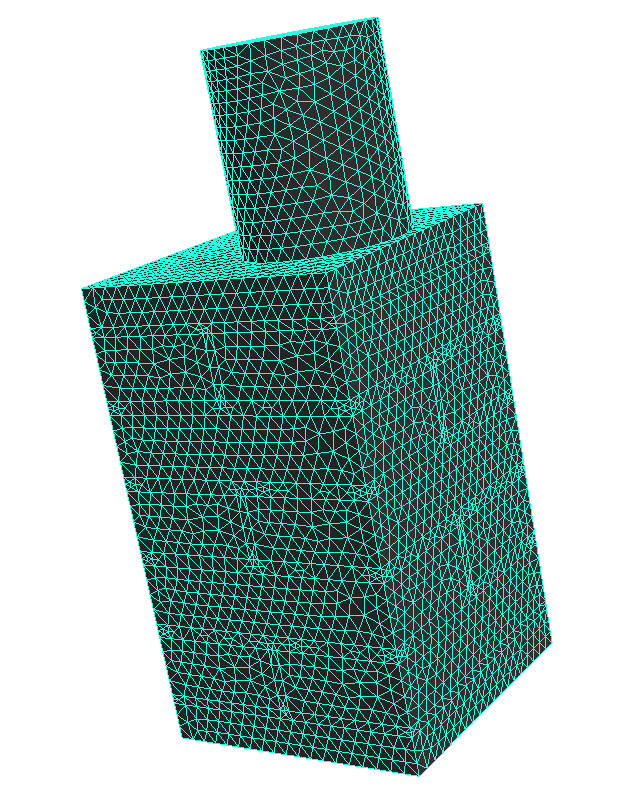}
\caption{FE mesh of the investigated masonry piers.}
\label{fig:piersMesh}
\end{figure}

The geometry of the 3D FE model was following the geometry of the experimentally tested masonry piers, as described in Figure~\ref{fig:schemePiers}; the FE mesh is presented in Figure~\ref{fig:piersMesh}. The interface between bricks and surrounding mortar was not explicitly defined and modeled using interface elements, because the interface was not the weakest link in tension, recall Section~\ref{sec:parametersAcquisition}.

%The material model input data were obtained from the inverse analysis, as described in Section~\ref{sec:inverseAnalysis}, while the auxiliary steel plates were modeled as an isotropic elastic continuum. Despite the availability of interface strength testing results, the interface between bricks and surrounding mortar was not explicitly defined and modeled using interface elements, since their introduction did not have any impact on results while the computational costs were rather big.

In order to define the loading of the piers and boundary conditions in a realistic way, the model also consisted, beside the auxiliary steel slabs, of the cylindrical load cell, both modeled as an isotropic elastic continuum. The loading was accomplished by an incremental displacement imposed to a single node at the crest of the steel cylinder in order to allow rotations around all axes. The load-step increments were adjusted for each loading stage individually in order to reach convergence for a minimum computational cost.

\begin{figure}[htp]
\centering
\includegraphics[width=0.9\textwidth]{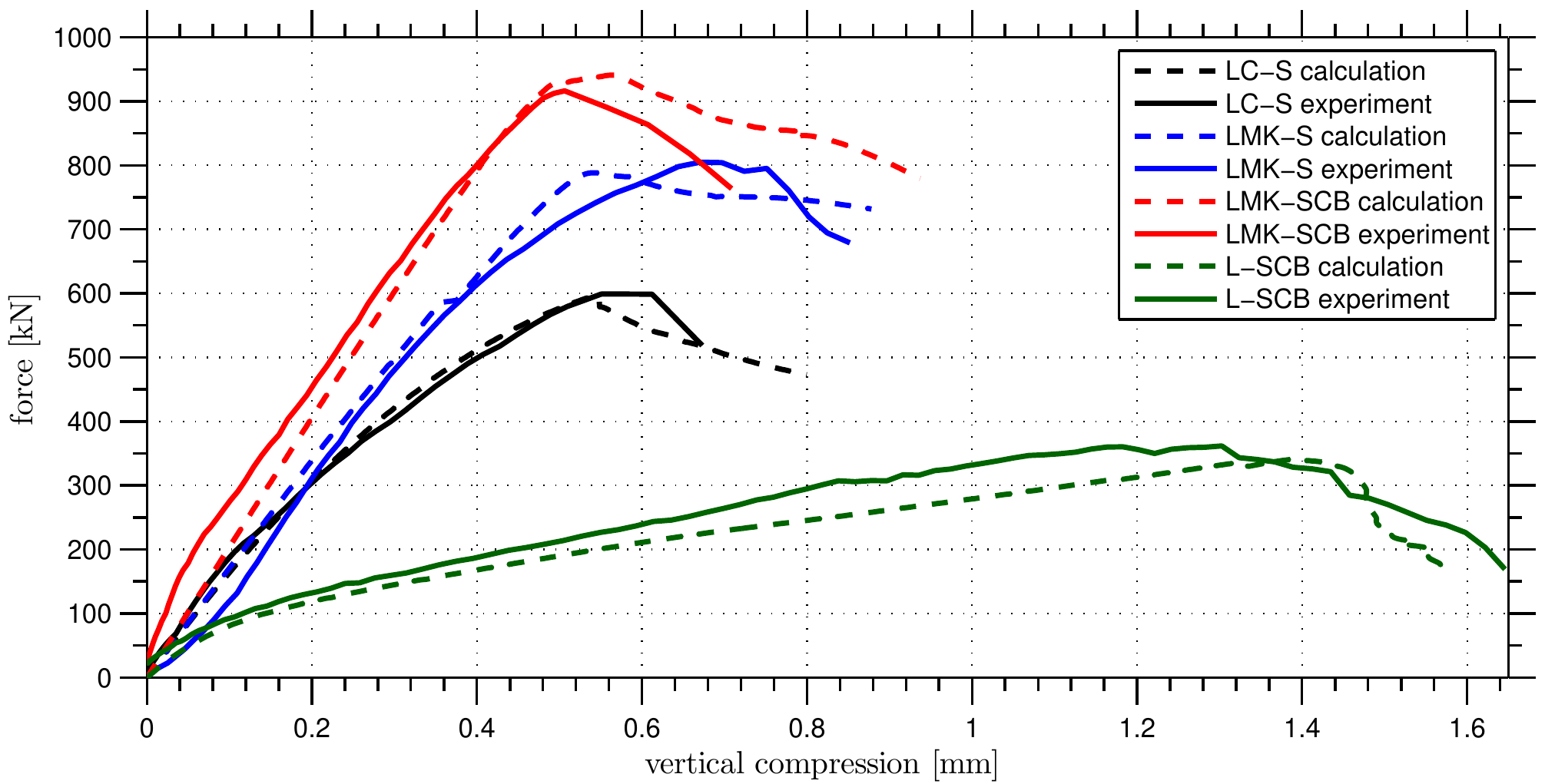}
\caption{Comparison between the numerically and experimentally obtained load-displacement diagrams, describing the response of the tested masonry piers to the imposed vertical displacement at the crest of the loading cell.}
\label{fig:diagramsPiers_vertical}
\end{figure}

The FE model was at first validated by comparing the predicted and measured load-displacement diagrams, as presented in Figure~\ref{fig:diagramsPiers_vertical}. Both, the reactions in a load cell and the displacements obtained from DIC by placing virtual extensometers at the top and bottom of the tested piers, were in a good agreement with the numerically obtained predictions, Figure~\ref{fig:schemePiers}. The agreement between the predicted cracking patterns (reflected in the field of damage distribution) and the observed development of cracks (visualized as a map of maximum principal strains obtained from DIC) is not perfect for all tested piers. However, considering the non-homogeneous microstructure of the tested materials, the FE analysis results cab be considered satisfactory.

%The agreement between the predicted crack patterns (reflected in the map of damage distribution) and cracking patterns resulting from the experimental destruction of the masonry piers (visualized as a map of maximum principal strains obtained from DIC) is not perfect for all tested piers, however, the results are satisfactory.

\begin{figure}[htp]
\centering
   \begin{subfigure}{0.43\linewidth} \centering
    \includegraphics[width=\textwidth]{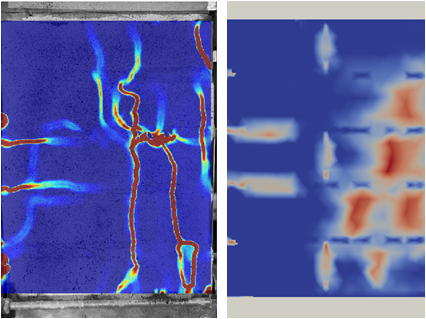}
    \caption{LC-S mortar.}
    \label{fig:crackPattern_LC-S}
   \end{subfigure}
   \hspace{0.1\linewidth}
   \begin{subfigure}{0.43\linewidth} \centering
    \includegraphics[width=\textwidth]{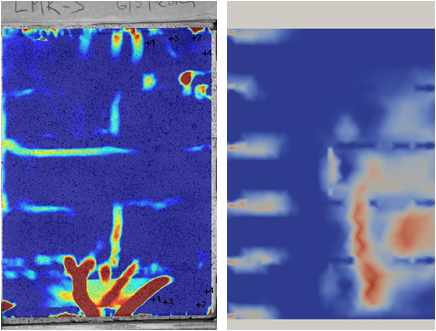}
    \caption{LMK-S mortar.}
    \label{fig:crackPattern_LMK-S}
   \end{subfigure}
   \vspace{5pt}
   \begin{subfigure}{0.43\linewidth} \centering
    \includegraphics[width=\textwidth]{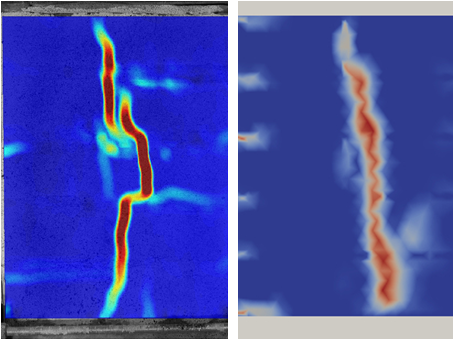}
    \caption{LMK-SCB mortar.}
    \label{fig:crackPattern_LMK-SCB}
   \end{subfigure}
   \hspace{0.1\linewidth}
   \begin{subfigure}{0.43\linewidth} \centering
    \includegraphics[width=\textwidth]{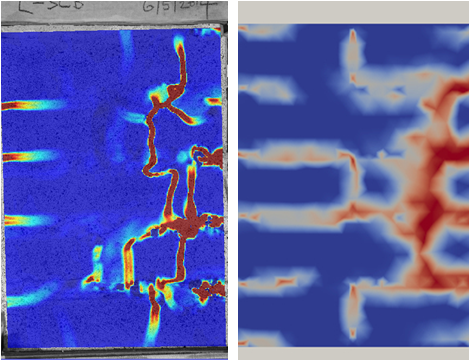}
    \caption{L-SCB mortar.}
    \label{fig:crackPattern_L-SCB}
   \end{subfigure}
\caption{Field of maximum principal strain obtained by DIC (left within the couples) and FE model damage predictions (right) on the face of the tested piers after reaching the peak load. The images provide a qualitative comparison between the predicted cracking patterns and experimental observations using DIC; the red areas represent the localized cracks.}
\label{fig:calibrationAllPiers}
\end{figure}

In the case of piers with lime-cement mortar LC-S (Figure~\ref{fig:crackPattern_LC-S}), the model correctly predicted the formation of multiple cracks at the compressed side of the tested piers and the formation of two major cracks at the opposite side due to tensile stresses from the pier bending. The DIC results in the case of lime-metakaolin mortar LMK-S (Figure~\ref{fig:crackPattern_LMK-S}) were influenced by a spalling of pier surface at the bottom, but the major crack formation in the middle of the tested pier can be identified in both, model predictions and DIC results. On the other hand, the formation of the major splitting crack in the case of mortar LMK-SCB (Figure~\ref{fig:crackPattern_LMK-SCB}), containing metakaolin and crushed bricks, was perfectly predicted by the FE simulation, as well as the distributed cracking at the compressed pier edge in the case of the weak mortar L-SCB (Figure~\ref{fig:crackPattern_L-SCB}).

In conclusion, the strategy to model the masonry units and mortar separately allowed us to capture the failure mode quite realistically (see Figure~\ref{fig:calibrationAllPiers}), enabling to study the relationship between the mechanical resistance of the masonry piers and bed-joint mortar properties.

\subsection{Influence of Mortar Properties on Mechanical Resistance of Masonry Piers} \label{sec:caseStudy}
The aim of the presented study was to show the relationship between the individual mortar material parameters and the load-bearing capacity of masonry piers having the same configuration of geometry and loading conditions as described in Sections~\ref{sec:experimentalTesting} and~\ref{sec:numericalSimulationsOfPiers}. The lime-metakaolin mortar without crushed bricks (LMK-S) was chosen as the reference material, for which a single material parameter was changed at a time to assess its impact on the load-bearing capacity of the masonry pier.

Such analysis clearly indicated what the key material parameters were, and how to optimize the mortar composition towards a higher mechanical resistance of masonry structures. Similar approach was adopted e.g.~by Sandoval and Roca~\cite{Sandoval_2012}, who studied the influence of geometry and material properties of individual constituents on the buckling behavior of masonry walls.

\subsubsection{Influence of Mortar Stiffness}

\begin{figure}[htp]
\begin{minipage}[b]{0.48\linewidth}
\centering
    \includegraphics[width=\textwidth]{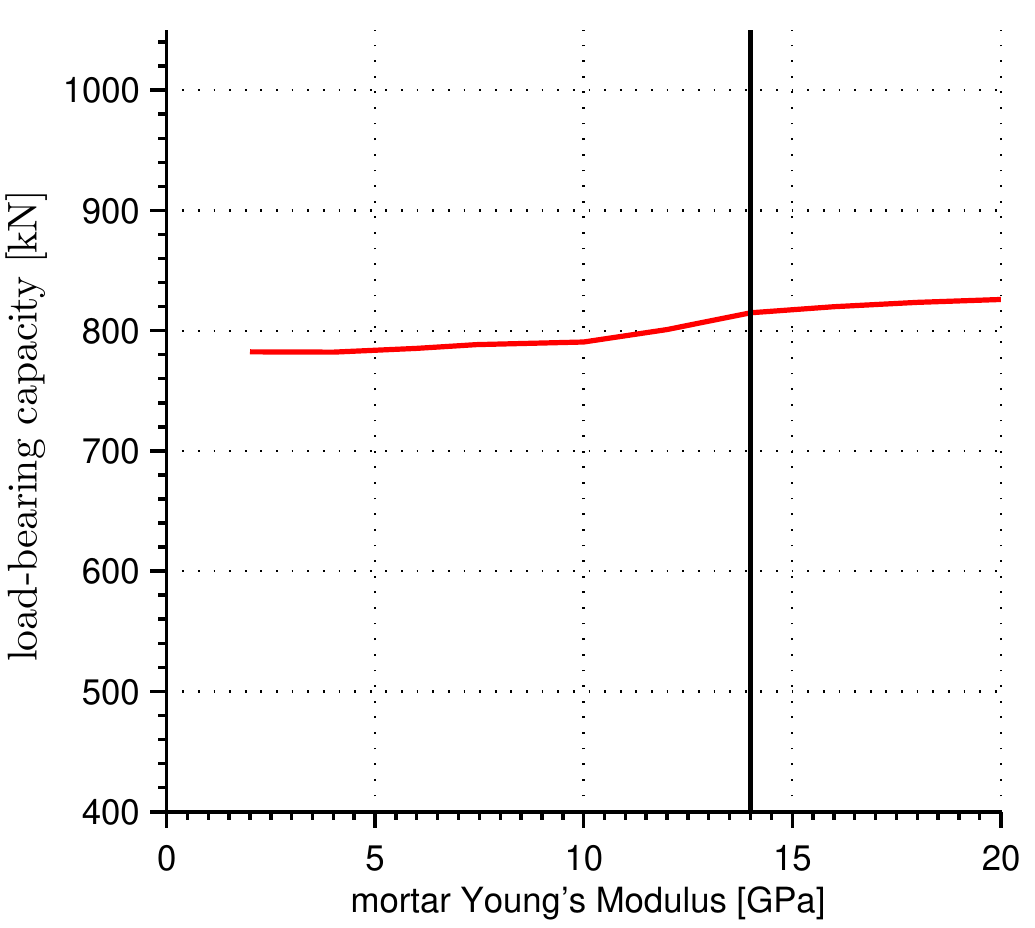}
    \caption{Dependence of the masonry pier maximum load-bearing capacity on mortar stiffness; black line represents the brick Young's modulus $E^{\mathrm{(b)}}$~=~14~GPa.}
    \label{fig:dependence_E}
\end{minipage}
\hspace{0.3cm}
\begin{minipage}[b]{0.5\linewidth}
\centering
    \begin{subfigure}{0.48\linewidth} \centering
     \includegraphics[width=\textwidth]{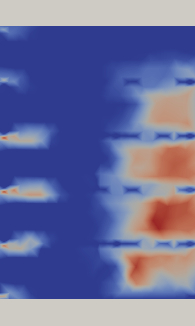}
   \end{subfigure}
   \hspace{0.01\linewidth}
   \begin{subfigure}{0.48\linewidth} \centering
     \includegraphics[width=\textwidth]{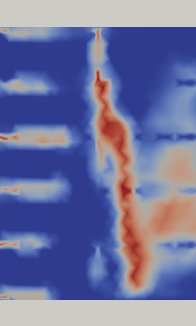}
   \end{subfigure}
   \caption{Crack patterns on masonry piers containing compliant ($E^{\mathrm{(m)}}$~=~2~GPa, left) and stiff ($E^{\mathrm{(m)}}$~=~20~GPa, right) mortar.}
   \label{fig:crackPatterns_E}
\end{minipage}
\end{figure}

The plot in Figure~\ref{fig:dependence_E} clearly demonstrates that the value of the mortar Young's modulus has just a minor impact on the load-bearing capacity of the studied masonry pier, and that there is no abrupt change when the mortar stiffness becomes superior to the stiffness of masonry units. However, the failure mode changes quite significantly. The use of a compliant mortar results in a multiple cracking of bricks at the more loaded side due to poor supporting, while a major crack passing through the entire column in the middle forms if the bed joints are stiff, see Figure~\ref{fig:crackPatterns_E}.

It could seem advantageous to use mortars lacking pozzolanic additives because of their lower stiffness, in order to produce masonry of a higher deformation capacity within the elastic range. Such masonry would in theory better resist seismic loading or imposed displacements, e.g. due to differential subsoil settlement. However, the compliant pure-lime mortars without additives promoting the hydraulic reactions are weak and suffer from an increased shrinkage cracking~\cite{Nezerka_2013_pastes}.

\subsubsection{Influence of Mortar Tensile Strength}

The tensile strength and fracture energy had to be modified carefully at the same time in order to avoid snap-back in the stress-strain diagram, and to preserve the same post-peak ductility for all investigated mortars.

\begin{figure}[htp]
\begin{minipage}[b]{0.48\linewidth}
\centering
    \includegraphics[width=\textwidth]{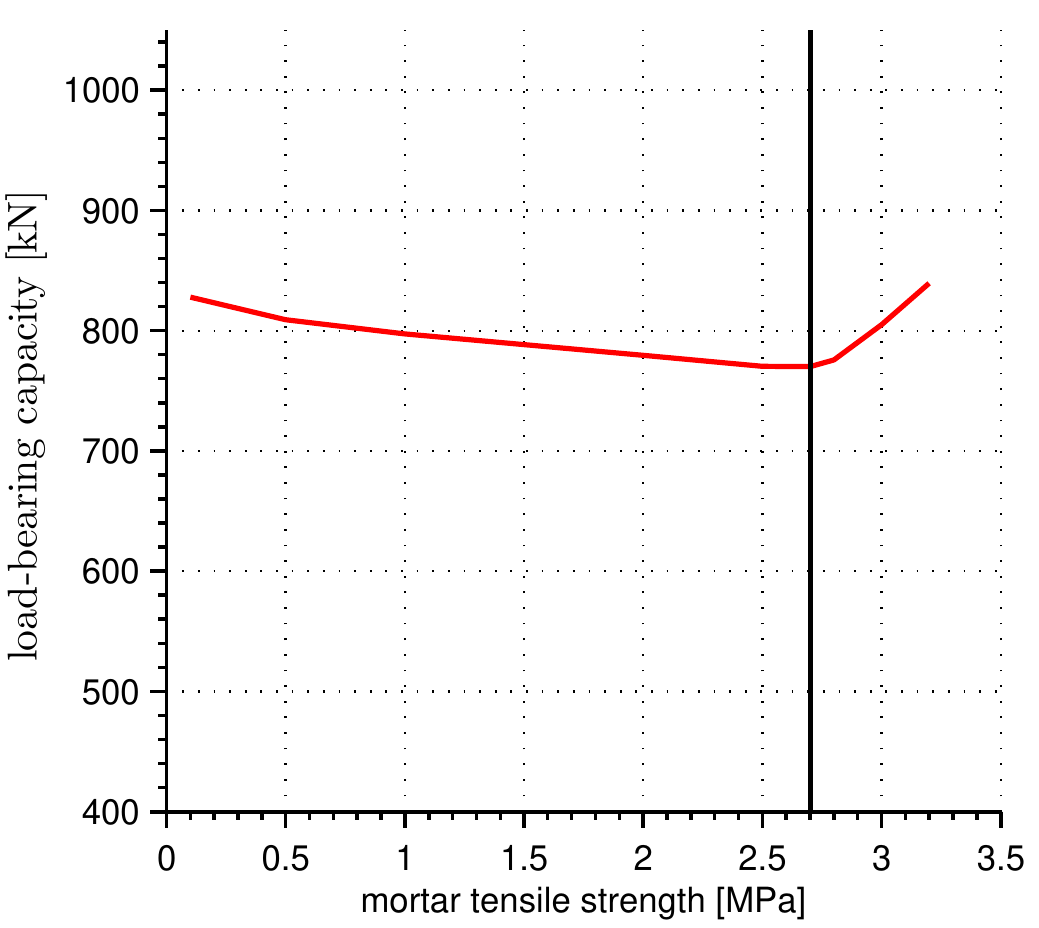}
    \caption{Dependence of the masonry pier maximum load-bearing capacity on mortar tensile strength; black line represents the brick tensile strength $f_{\mathrm{t}}^{\mathrm{(b)}}$~=~2.7~MPa.}
    \label{fig:dependence_ft}
\end{minipage}
\hspace{0.3cm}
\begin{minipage}[b]{0.5\linewidth}
\centering
    \begin{subfigure}{0.48\linewidth} \centering
     \includegraphics[width=\textwidth]{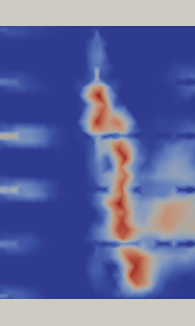}
   \end{subfigure}
   \hspace{0.01\linewidth}
   \begin{subfigure}{0.48\linewidth} \centering
     \includegraphics[width=\textwidth]{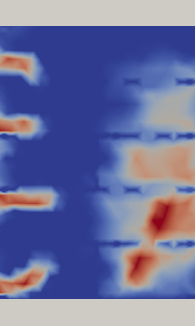}
   \end{subfigure}
   \caption{Crack patterns on masonry piers containing weak ($f_{\mathrm{t}}$~=~0.1~MPa, left) and strong ($f_{\mathrm{t}}$~=~3.2~MPa, right) mortar in tension.}
   \label{fig:crackPatterns_ft}
\end{minipage}
\end{figure}

Given the studied masonry pier and the boundary conditions, the tensile strength appears to have just a minor effect if it is lower than the strength of masonry units (bricks), see Figure~\ref{fig:dependence_ft}. On the other hand, the mortars of higher strength in tension act as a confinement of the eunits and the masonry reinforcement. Since common lime- or cement-based mortars hardly attain the tensile strength superior to the strength of masonry units, the bed joint strengthening is accomplished e.g.~by means of embedded steel rods~\cite{Valluzzi_2005}.

According to our numerical simulations, the confinement imposed by the strong mortars resulted in the cracking of the bricks and eventually the formation of the wedge-like failure as opposed to the vertical splitting of the pier containing a very weak mortar as indicated in Figure~\ref{fig:crackPatterns_ft}.

\subsubsection{Influence of Mortar Compressive Strength}

The bed-joint mortar compressive strength appears to be the crucial parameter with respect to the load-bearing capacity of masonry subjected to a combination of compression and bending. Mortars of a low compressive strength suffer an irreversible deformation at relatively low levels of external load, and masonry units are consequently subjected to uneven distribution of stresses due to imperfect supporting and excessive deformation of the bed joints.

\begin{figure}[htp]
\begin{minipage}[b]{0.48\linewidth}
\centering
    \includegraphics[width=\textwidth]{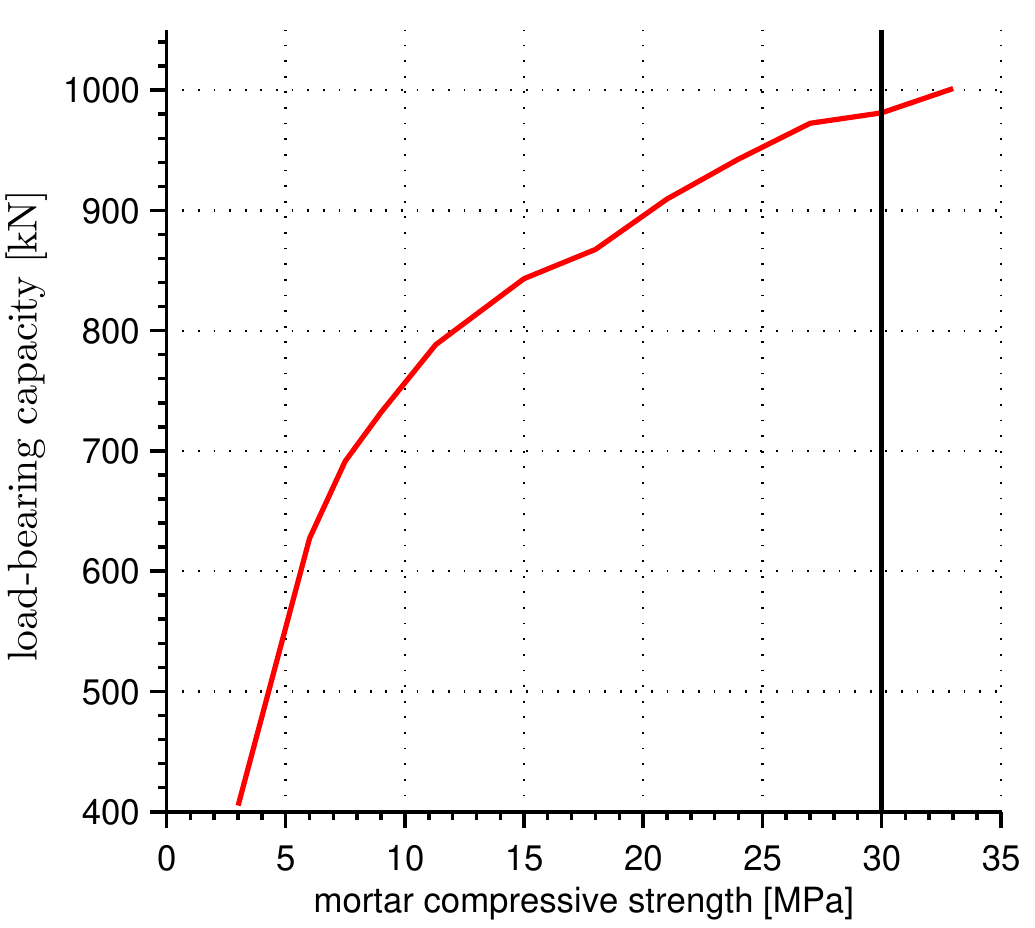}
    \caption{Dependence of the masonry pier maximum load-bearing capacity on mortar compressive strength; black line represents the brick compressive strength $f_{\mathrm{c}}^{\mathrm{(b)}}$~=~30~MPa.}
    \label{fig:dependence_fc}
\end{minipage}
\hspace{0.3cm}
\begin{minipage}[b]{0.5\linewidth}
\centering
    \begin{subfigure}{0.48\linewidth} \centering
     \includegraphics[width=\textwidth]{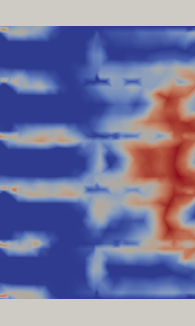}
   \end{subfigure}
   \hspace{0.01\linewidth}
   \begin{subfigure}{0.48\linewidth} \centering
     \includegraphics[width=\textwidth]{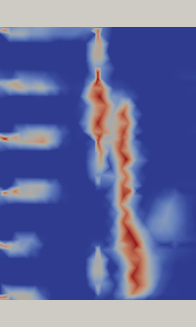}
   \end{subfigure}
   \caption{Crack patterns on masonry piers containing weak ($f_{\mathrm{c}}$~=~3~MPa, left) and strong ($f_{\mathrm{c}}$~=~33~MPa, right) mortar in compression.}
   \label{fig:crackPatterns_fc}
\end{minipage}
\end{figure}

In the case of the modeled masonry piers, the early crushing of the weak bed-joint mortar resulted in cracking of bricks at the more loaded pier periphery, Figure~\ref{fig:crackPatterns_fc}. This phenomenon limited the load-bearing capacity of the tested pier rather significantly, especially in the case of very poor mortars ($f_{\mathrm{c}}^{\mathrm{(m)}}<10$~MPa), see Figure~\ref{fig:dependence_fc}. The bed joints containing mortars of a high compressive strength did not suffer the inelastic deformation before the major splitting vertical crack appeared due to transversal expansion, resulting in a high load-bearing capacity. Therefore, the mortars with superior compressive strength should be used especially if a bed joint reinforcement is introduced so that the high strength can be efficiently exploited.

\section{Discussion of Results} \label{sec:resultsDiscussion}

The eccentrically compressed masonry pier was selected as a model example to address both, behavior in compression, being the most frequent loading of masonry elements, and tension, which is considered critical for masonry. The performance of the conventionally used lime-cement mortar was compared with mortars containing the pozzolanic alternative --- metakaolin. To reach even better performance, crushed brick fragments were also used to replace a portion of stiff river sand. Such approach was adopted based on findings from the previous studies, e.g.~\cite{Nezerka_2013_pastes, Nezerka_2014_Brno, Nezerka_2014_EAN2013}, claiming that mortars containing active pozzolans and relatively compliant crushed brick fragments exhibit a superior strength.

Series of compression and three-point bending tests were carried out primarily in order to obtain the input parameters characterizing individual materials in the FE model. The results of the basic material tests conclusively indicate that the addition of metakaolin provides the lime-based mortars with significantly higher strength than the addition of Portland cement. On the other hand, the pure-lime mortars lacking any additives appeared to be very poor. These findings are in agreement with several studies; e.g. by Vejmelkov\'{a} et al.~\cite{Vejmelkova_2012} claiming that by replacing 20~\% of lime with metakaolin the mortar compressive strength can increase up to five times and the flexural strength up to three times, which is in agreement with the study by Velosa et al.~\cite{Velosa_2009}.

The partial replacement of sand grains by crushed brick fragments further improved the mechanical performance of the mortars, justifying their extensive use in ancient times~\cite{Mallinson_1987, Degryse_2002, Baronio_1997}. The higher strength is attributed to a reduction of shrinkage-induced cracking due to presence of the compliant brick fragments. This in turn leads to a better mortar integrity as suggested by Ne\v{z}erka~et~al.~\cite{Nezerka_2014_nanoindentation, Nezerka_2013_pastes}, and lower stress concentrations in the vicinity of aggregates, as studied in detail in~\cite{Nezerka_2016_model}.

The superior strength of the metakaolin-enriched mortars was also reflected by the increased load-bearing capacity of the tested piers, in particular from 360~kN and 600~kN, reached in the case of pure-lime (L-SCB) and lime-cement mortar (LC-S), respectively, up to 800~kN when 30~\% of the binder was replaced by metakaolin (LMK-S). Moreover, the load-bearing capacity was further increased with the use of mortar containing crushed bricks (LMK-SCB), reaching up to 915~kN. This strength enhancement can explain the resistance and longevity of numerous ancient masonry structures containing cocciopesto~\cite{Moropoulou_2002, Moropoulou_2000}. The extraordinary strength of the LMK-SCB mortar together with the good adhesion between the mortar and bricks should also result in an increased seismic resistance, as suggested by Costa et al.~\cite{Costa_2012}.

Knowing the basic material parameters, the damage-plastic material model used for the 3D~FE simulations allowed to reproduce the experimental results with a relatively high accuracy, despite the complex composite action taking place in masonry. Even the simplest case of uniaxial compression leads to a triaxial compression in mortar, while introducing a uniaxial compression and biaxial tension in usually stiffer masonry units. Such scenario usually leads to the formation of vertical splitting cracks leading to a complete failure~\cite{Kaushik_2007}.

The chosen strategy to model the bricks and mortars as two distinct materials allowed to investigate the relationship between the individual material parameters and structural behavior of the masonry pier. Our findings that the mortar compressive strength has the biggest impact on the load-bearing capacity is in contradiction with the conclusions of Gumaste~et~al.~\cite{Gumaste_2007} and Pav\'{i}a and Hanley~\cite{Pavia_2010}. They claim that mortar compressive strength has just a minor impact on the behavior of masonry subjected to uniaxial compression. This discrepancy can be probably attributed to a different experimental set-up, in particular to the eccentricity of loading introduced in our study. The eccentric loading was responsible for a significant deformation of the bed joints, leading to a non-linear response at relatively early loading-stage.

The assumption that the difference between the Young's modulus of bricks and mortar is the precursor of the compression failure~\cite{Zucchini_2007} was not confirmed, and the load-bearing capacity of the masonry piers was almost independent of mortar stiffness.

\section{Conclusions} \label{sec:conclusions}

The chosen strategy to combine the comprehensive experimental analysis together with the numerical modeling revealed new findings to be considered during the design of bed joint mortars. Even though the study was focused purely on the lime-based mortars, because these are accepted by the authorities for cultural heritage, our findings can also help with the design of mortars and masonry based on modern materials.

The results of the basic material tests demonstrate the superior strength of mortars containing metakaolin, when compared to a pure-lime or lime-cement ones. The mortar strength was further increased by the addition of crushed bricks, which is attributed to the reduction of microcracking due to shrinkage around the relatively compliant ceramic fragments. It can be also conjectured that the hydraulic reaction in mortars containing metakaolin was promoted by the presence of water retained within the crushed brick fragments.

The enhanced strength of the metakaolin-rich mortars, and especially those containing crushed bricks, was reflected in the significantly increased load-bearing capacity of masonry piers loaded by the combination of compression and bending. This can explain the extraordinary resistance and durability of ancient masonry structures with cocciopesto mortars. Moreover, the utilization of the waste by-products from ceramic plants makes the material sustainable for a relatively low cost, since the fragments partially replace binder, being the most expensive mortar component.

Based on experimental observations the damage-plastic material model seemed to be the most appropriate to describe the constitutive behavior of mortars and bricks in the FE model. The chosen strategy to model the mortars and bricks as distinct materials allowed the relatively accurate reproduction of the experimentally obtained data in terms of the predicted crack patterns and load-displacement diagrams. Results of the numerical simulations and DIC analysis clearly demonstrate that the mortar properties have an enormous impact on the load-bearing capacity of masonry, strain localization, and the formation of cracks.

The numerical analysis, based on the FE model verified through the comprehensive experimental analysis, revealed that mortar compressive strength is the key material parameter with respect to the load-bearing capacity of the piers subjected to the combination of bending and compression. Considering the studied geometry and boundary conditions, tensile strength and mortar Young's modulus influence the pier behavior and modes of failure, however, do not have any significant impact on the load-bearing capacity.

\section*{Acknowledgments}
The authors acknowledge financial support provided the Czech Science Foundation, project No.~GA13-15175S, and by the Ministry of Culture of the Czech Republic, project No.~DF11P01OVV008.

\end{document}

%% file: figure03.pdf_tex
%% Creator: Inkscape 0.48.5, www.inkscape.org
%% Accompanies image file 'specimenScheme.pdf' (pdf, eps, ps)
\begingroup%
  \makeatletter%
  \providecommand\color[2][]{%
    \errmessage{(Inkscape) Color is used for the text in Inkscape, but the package 'color.sty' is not loaded}%
    \renewcommand\color[2][]{}%
  }%
  \providecommand\transparent[1]{%
    \errmessage{(Inkscape) Transparency is used (non-zero) for the text in Inkscape, but the package 'transparent.sty' is not loaded}%
    \renewcommand\transparent[1]{}%
  }%
  \providecommand\rotatebox[2]{#2}%
  \ifx\svgwidth\undefined%
    \setlength{\unitlength}{438.19063907bp}%
    \ifx\svgscale\undefined%
      \relax%
    \else%
      \setlength{\unitlength}{\unitlength * \real{\svgscale}}%
    \fi%
  \else%
    \setlength{\unitlength}{\svgwidth}%
  \fi%
  \global\let\svgwidth\undefined%
  \global\let\svgscale\undefined%
  \makeatother%
  \begin{picture}(1,1.29397624)%
    \put(0,0){\includegraphics[width=\unitlength]{figure03.pdf}}%
    \put(0.3980673,0.09473538){\color[rgb]{0,0,0}\makebox(0,0)[lt]{\begin{minipage}{0.26132123\unitlength}\raggedright 290\end{minipage}}}%
    \put(0.89167615,0.56406417){\makebox(0,0)[lt]{\begin{minipage}{0.17949507\unitlength}\raggedright Ext-1\end{minipage}}}%
    \put(0.4555949,1.28284843){\color[rgb]{0,0,0}\makebox(0,0)[lt]{\begin{minipage}{0.26132123\unitlength}\raggedright 50\end{minipage}}}%
    \put(0.55356863,1.29840766){\color[rgb]{0,0,0}\makebox(0,0)[lt]{\begin{minipage}{0.4424352\unitlength}\raggedright $F_{\mathrm{ext}}$\end{minipage}}}%
    \put(0.65221029,1.11211903){\color[rgb]{0,0,0}\makebox(0,0)[lt]{\begin{minipage}{0.26132123\unitlength}\raggedright 145\end{minipage}}}%
    \put(0.77485547,0.0719275){\color[rgb]{0,0,0}\makebox(0,0)[lt]{\begin{minipage}{0.26132123\unitlength}\raggedright [mm]\end{minipage}}}%
    \put(-0.00681272,0.56699373){\color[rgb]{0,0,0}\rotatebox{90}{\makebox(0,0)[lt]{\begin{minipage}{0.07510264\unitlength}\raggedright 385\end{minipage}}}}%
    \put(0.84854309,1.01685327){\color[rgb]{0,0,0}\rotatebox{90}{\makebox(0,0)[lt]{\begin{minipage}{0.09270859\unitlength}\raggedright 30\end{minipage}}}}%
    \put(0.84872329,0.25745432){\color[rgb]{0,0,0}\rotatebox{90}{\makebox(0,0)[lt]{\begin{minipage}{0.09270859\unitlength}\raggedright 30\end{minipage}}}}%
  \end{picture}%
\endgroup%